# Tailored tunnel magnetoresistance response in three ultrathin chromium trihalides

Hyun Ho Kim[1*], Bowen Yang[1], Shangjie Tian[2], Chenghe Li[2], Guo-Xing Miao[1], Hechang Lei[2], and Adam W. Tsen[1*]

[1]*Institute for Quantum Computing, Department of Chemistry, Department of Physics and Astronomy, and Department of Electrical and Computer Engineering, University of Waterloo, Waterloo, Ontario N2L 3G1, Canada*

[2]*Department of Physics and Beijing Key Laboratory of Opto-electronic Functional Materials & Micro-Nano Devices, Renmin University of China, Beijing 100872, China*

[*]Correspondence to: hyun.kim@uwaterloo.ca, awtsen@uwaterloo.ca

## Abstract

Materials that demonstrate large magnetoresistance have attracted significant interest for many decades. Recently, extremely large tunnel magnetoresistance (TMR) has been reported by several groups across ultrathin $CrI_3$ by exploiting the weak antiferromagnetic coupling between adjacent layers. Here, we report a comparative study of TMR in all three chromium trihalides ($CrX_3$, X= Cl, Br, or I) in the two-dimensional limit. As the materials exhibit different transition temperatures and interlayer magnetic ordering in the ground state, tunneling measurements allow for an easy determination of the field–temperature phase diagram for the three systems. By changing sample thickness and biasing conditions, we then demonstrate how to maximize and further tailor the TMR response at different temperatures for each material. In particular, near the magnetic transition temperature, TMR is non-saturating up to the highest fields measured for all three compounds owing to the large, field-induced exchange coupling.

Materials that show a substantial change in resistance in response to an applied magnetic field, or large magnetoresistance (MR), are generally rare but highly sought-after, both for fundamental interest as well as for potential applications in magnetic sensors and memory[1-6]. In non-magnetic semimetals, such as $WTe_2$[3] or $MoTe_2$[7], extreme *positive* MR ($10^6$% at 10T) can be achieved at low temperature due to both a balance of electron and hole carriers[8-10] as well as their relatively high mobilities[11, 12]. The MR is further non-saturating up to the highest fields measured. In magnetic thin-films or perovskite manganites, respectively, giant ($10^3$%)[13] or colossal ($10^6$%)[14] *negative* MR can instead be achieved at higher temperatures. The former effect forms the basis of magnetic memory technology due to the low critical fields needed to saturate the MR.

The recent discoveries of 2D magnets[15-17], such as $CrI_3$, provide a novel opportunity to explore large MR materials through the use of van der Waals heterostructures. Despite being insulating, electrons can tunnel through the $CrI_3$ layers in the ultrathin limit. Several groups have previously demonstrated negative tunnel magnetoresistance (TMR) as large as $10^6$% in graphene/$CrI_3$/graphene junctions under a 2T field at optimized biasing voltage[18-20]. Here, as the antiferromagnetically coupled $CrI_3$ layers become polarized by the magnetic field[21-23], the tunneling current increases dramatically due to an electron spin-filtering effect[24]. This result motivates a systematic study of TMR in the entire $CrX_3$ (X= I, Br, and Cl) family across bias, temperature, sample thickness, as well as contact material. For $CrI_3$ and $CrCl_3$ with antiferromagnetic (AFM) interlayer coupling, we observe the largest TMR in slightly thicker samples at low temperature, but which saturates at the critical polarizing field of 2T, while $CrBr_3$ samples with ferromagnetic (FM) interlayer coupling always show negligible TMR in the ground state. Near the transition temperature, however, all three materials demonstrate non-saturating TMR up to 14T. We interpret this effect as a field-induced splitting of the paramagnetic electron bandstructure. Surprisingly, this splitting can be over 80% of the value in the ground state. Finally, using NbSe$_2$ with larger work function in place of graphene, we find that exchange splitting is much smaller for hole carriers in the valence band, consistent with theoretical expectations. Our results establish CrX$_3$ as a highly tunable platform to study TMR physics using relatively simple device geometries.

A schematic illustration of our devices is shown in the left panel of Fig. 1a and the detailed fabrication procedure can be found in the Methods section. In short, ultrathin CrI$_3$, CrBr$_3$, or CrCl$_3$ from 2 to 15 layers (L) is exfoliated in a nitrogen-filled glovebox and sandwiched between top and bottom few-layer graphene electrodes (Gr) with full encapsulation by hexagonal boron nitride (hBN) on both sides for sample protection. In the right panel of Fig. 1a, we show an optical image of such a CrI$_3$ device as an example. Our key findings can be summarized as a plot of TMR vs. thickness for each material measured at their "optimal" temperature, as shown in Fig. 1b. While their differences shall be explained in detail in the remaining sections, we note that the largest TMR response is generally observed in samples over 6L, and so we show main measurements on devices in this thickness regime.

We begin with transport behavior without magnetic field in order to confirm the ground-state magnetic properties of CrX$_3$. In Fig. 2, we show temperature-dependent I-V characteristics for few-layer devices incorporating the three different materials (CrI$_3$: 8L, CrBr$_3$: 8L, and CrCl$_3$: 15L). All tunnel junctions exhibit nonlinear current-voltage characteristics due to quantum mechanical tunneling across the insulating layers. At higher bias yielding current above ~1nA, transport is governed by Fowler-Nordheim (FN) tunneling[18], while for direct tunneling near zero bias, the current is below the noise level (~1pA) for samples of this thickness. A zero-bias conductivity is measurable in thinner samples, however (see Supplementary Fig. S1). For CrI$_3$ and

$CrCl_3$, decreasing temperature lowers the current level for a given voltage, whereas $CrBr_3$ shows the opposite trend. In order to verify this, we additionally measured resistance with continuously changing temperature, and the results are shown in the corresponding insets in the bottom panel of Fig. 2. Here, a marked kink (or peak) was observed at the magnetic transition temperature ($T_{AFM}$ or $T_{FM}$) for each device. By taking the derivative of these curves, we determined the critical temperatures more precisely to be 46K for $CrI_3$, 37K for $CrBr_3$, and 17K for $CrCl_3$ (see Supplementary Fig. S2). These temperatures are consistent with previous reports for bulk $CrBr_3$[25] and $CrCl_3$[26], although few-layer $CrI_3$ shows a slightly lower value than that for the bulk crystal (61K)[27], possibly due to different interlayer spin coupling and/or stacking between the two systems[28, 29]. The abrupt change in tunnel resistance at the critical temperature arises from spontaneous exchange splitting of the $CrX_3$ conduction band[30, 31]. We shall later discuss the results for the valence band. Below $T_{AFM}$ ($T_{FM}$), increasing (decreasing) resistance with decreasing temperature for $CrI_3$ and $CrCl_3$ ($CrBr_3$) indicates that the ground-state interlayer magnetic ordering is AFM (FM) between adjacent layers, as tunneling across filters with antiparallel (parallel) spins raises (lowers) the effective barrier height. Schematic illustrations of these effects are shown in the top panel of Fig. 2. These results are consistent with several recent studies of ultrathin $CrX_3$[32-34].

The different ground-state interlayer spin ordering observed in these three materials will yield different phase transitions and TMR behavior upon application of a magnetic field, both of which can be measured in our device geometry. In Fig. 3a-c, we show voltage vs. temperature at constant current for the same three devices at several different fields applied perpendicular to the layers. We chose a small, 0.1nA current level in order to minimize the effects of Joule heating. Nevertheless, we found that measured transition temperatures did not change up to 100nA (Supplementary Fig. S3). In general, increasing magnetic field polarizes spins parallel to the field direction. For CrI$_3$ and CrCl$_3$, the field decreases the resistance at low temperature as the interlayer AFM ground state is gradually destroyed, consistent with magneto-optical studies[16, 21-23]. For CrBr$_3$, however, the low-temperature resistance is barely changed, since a spin-parallel, interlayer FM state has already formed in the ground state. Instead, the largest change occurs around the critical temperature. Here, increasing magnetic field stabilizes the parallel spin state, and so the voltage is decreased while the transition temperature into this state, $T_P$, is increased. In CrI$_3$ and CrCl$_3$, we observe that for intermediate magnetic fields, not only is $T_P$ pushed to higher temperatures, but a second transition to an antiparallel interlayer spin state can be seen at temperatures below $T_{AP}$ and decreases for increasing field. In the limit of zero field, $T_P$ and $T_{AP}$ approach $T_{FM}$ and $T_{AFM}$, respectively, critical temperatures for the transition to *spontaneous* magnetic order. These net results are reproducible over several different samples (see Supplementary Fig. S4). We have additionally performed field-sweep measurements at several different temperatures (see Supplementary Fig. S5). The combined dataset allows us to obtain a field–temperature phase diagram for all three materials, as shown in Fig. 3d-f. In the spin schematics for the ground state, the easy axis is drawn to be out of plane for CrI$_3$ and CrBr$_3$, while it is in-plane for CrCl$_3$, as had been just recently demonstrated[32-34].

A field-induced transition to the spin-parallel state should be accompanied by substantial TMR. We show I-V plots with and without a 5.5T perpendicular field for CrI$_3$ and CrCl$_3$ (insets of Fig 4a and b) and for CrBr$_3$ (see inset of Supplementary Fig. S6a) at 1.4K. When the field is applied, we generally observe clear TMR as the current is enhanced for any given voltage. For constant voltage biasing $V$, we define the TMR percentage between 0 and 5.5T as TMR (%) = $\frac{I(5.5T,V) - I(0T,V)}{I(0T,V)} \times 100\%$[24], and have plotted these voltage-dependent values in the main panels of Fig. 4a, Fig. 4b, and Fig. S6a for CrI$_3$, CrCl$_3$, and CrBr$_3$, respectively. First, we notice that the devices exhibit a noticeable asymmetry between negative and positive voltage. While we are unclear as to the precise origin of this effect, we note that there is an inherent asymmetry built into the device geometry as the top and bottom graphite layers are not identical. Different doping levels between the two will lead to different barrier heights and cause asymmetric I-V as our later analysis will show. Similar behavior was also observed in previous reports[18, 19]. Nevertheless, at an optimized voltage level, we find that TMR at 1.4K reaches as high as $6 \times 10^5$% in $CrI_3$ and 1490% in $CrCl_3$ (Fig. 4a and b), whereas a much smaller TMR of ~10% was observed in $CrBr_3$ (Fig. S6), owing to the pre-formed, spin-parallel ground state. Overall, we attribute the larger TMR values to the high-quality interfaces naturally formed between the crystalline layers, as well as the sizable spin splitting of the electron bandstructure, which we shall discuss in detail below.

In order to determine the smallest magnetic field necessary to achieve the large TMR, we measured normalized current at the optimal voltage with continuously changing field. These results are shown in main panel of Fig. 4c. We observe that the TMR response for $CrI_3$ and $CrCl_3$ effectively saturates at 2T, the critical polarizing field for these two materials in the ground state[32]. Up to this field, however, the current for $CrI_3$ increases in discrete steps, while that for $CrCl_3$ increases continuously. This could be understood by the difference in magnetic anisotropy between the two[31, 32]. Within the layers, $CrI_3$ exhibits the characteristics of a strong Ising ferromagnet with out-of-plane easy axis. A perpendicular field thus acts to flip the layers that are initially polarized in the opposite direction one by one. $CrCl_3$, on the other hand, has an in-plane easy axis, and so the same field smoothly rotates the spins until they are fully aligned. Finally, TMR in the $CrBr_3$ device is small and negligible up to 5.5T. We next repeated these measurements at many different temperatures and have plotted the TMR percentage for all three devices as a function of temperature in Fig. 4d. While both $CrI_3$ and $CrCl_3$ show decreasing TMR with increasing temperature, CrBr$_3$ shows the opposite trend up to 40K, at which its maximum TMR is found to be ~200% (see also Supplementary Fig. S6b). In particular, the values for CrBr$_3$ and CrCl$_3$ cross at ~30K.

The contrasting TMR behavior in the three materials could be quantitatively explained by considering spin-dependent tunneling in the FN regime. Here, the current-voltage relation scales as $\ln\left(\frac{I_{\uparrow\downarrow}}{V^2}\right) \sim \frac{\Phi_{\uparrow\downarrow}^{3/2}(B)}{V}$, where $\Phi_{\uparrow\downarrow}(B)$ is the tunnel barrier for electrons of different spin in CrX$_3$ and is tunable by magnetic field. As the work function of few-layer graphene[35] is expected to be close to the electron affinity of CrX$_3$[36], the tunneling current is dominated by electron carriers. For CrI$_3$ and CrCl$_3$ in the anti-parallel ground state, electrons of both spins encounter a spatially modulated barrier with effective height $\Phi_{\uparrow,eff}^{AP} = \Phi_{\downarrow,eff}^{AP}$. In the spin-parallel state with ≳ 2T upward field, the barrier becomes uniform, with $\Phi_\uparrow^P < \Phi_{eff}^{AP} < \Phi_\downarrow^P$. The lower barrier for spin-up electrons yields an exponential rise in spin-filtered current, which manifests as large TMR. A schematic bandstructure outlining these quantities is shown in the inset of Fig. 4c. The spin-degenerate barrier in the paramagnetic state near the transition temperature, $\Phi^{PM}$, is also shown, together with the exchange-induced gap, $E_{ex} = \Phi_\downarrow^P - \Phi_\uparrow^P = 2(\Phi^{PM} - \Phi_\uparrow^P)$. By fitting the current-voltage data to the FN formula at different magnetic field levels and temperatures, we are able to extract these various barrier heights. More detailed information about this process can be found in Supplementary Information: Section V. These quantities are summarized in Table 1 for all three compounds. In particular, CrI$_3$ shows larger TMR than CrCl$_3$ due to larger $E_{ex}$ (lower $\Phi_\uparrow^P$) relative to $\Phi_{eff}^{AP}$. The exchange splitting decreases with increasing temperature and gradually disappears above $T_{AFM}$[18].

In contrast, for CrBr$_3$ at low temperature, increasing field has a negligible effect on the barriers since they are already split uniformly in the P state. The peak TMR in CrBr$_3$ starting in the paramagnetic (PM) state near $T_{FM}$ ~ 40K then suggests a different mechanism. In order to explore this further, we have measured current vs. field at optimal voltage biasing for all three materials at their respective magnetic transition temperatures. In Fig. 5a, we show this dependence with current normalized to that at zero field. In all cases, current increases smoothly without saturation up to 14T, with $CrI_3$ showing the largest change, corresponding to a TMR value of 17000%.

At these temperatures, increasing magnetic field induces the splitting of $\Phi^{PM}$, with $\Phi^{P}_{\uparrow} < \Phi^{PM} < \Phi^{P}_{\downarrow}$, also resulting in considerable TMR. We have further extracted $E_{ex}$ as a function of magnetic field by performing full current-voltage measurements, and the results are plotted in Fig. 5b. For comparison, the respective exchange splitting in the ground state is marked as a dashed line. Relative to this value, at 14T the field-induced splitting at the transition temperature is already ~82% in $CrI_3$, ~85% in $CrBr_3$, and ~ 87% in $CrCl_3$, all two orders of magnitude larger than the Zeeman energy (~1.7meV at 14T). This effect may be analogous to that observed in the colossal MR materials undergoing a PM insulator to FM metal transition, wherein the field suppresses the resistance caused by critical spin fluctuations or phase disorder[14]. This surprising result indicates that large exchange coupling can be induced in these materials at elevated temperatures even with moderate magnetic fields.

Finally, all the analysis of the electron barrier heights so far was carried out on devices using few-layer graphene electrodes with relatively low work function (4.42eV)[35]. We have additionally investigated the spin splitting of the valence band in $CrI_3$ using few-layer $NbSe_2$ contacts with a higher work function (~5.9eV[37]). The results are discussed in Supplementary Information: Section VI. Overall, the critical temperature and field remain similar; however, TMR is ~$10^4$ times lower (~70%), from which we estimate $E_{ex}$ to be 2–3meV, consistent with theoretical expectations[31].

In summary, we have investigated the magnetic properties of all three chromium halides in the atomically thin limit by incorporating them in van der Waals tunnel junctions. We have systematically characterized both the ground state and field-driven phases as well as their TMR behavior with changing bias, temperature, thickness, and metal contact. While thicker samples with smaller work function graphene electrodes generally show higher TMR due to multiple spin filters of tunneling electrons acting in series, we find that the field range of the TMR response can be further tuned with temperature. Our work will have important implications for future devices utilizing these 2D materials.

# Methods

**Crystal synthesis.** CrI$_3$ and CrCl$_3$ single crystals were grown by the chemical vapor transport method. We used a two-zone horizontal tube furnace. The temperature for source (growth) zones was gradually raised to 993 – 873K (823 – 723K) within 24 hours, and then held for 150 hours for actual growth. CrBr$_3$ single crystals was purchased from HQ graphene.

**Device fabrication.** Graphite (CoorsTek), h-BN (HQ graphene), CrI$_3$, CrBr$_3$, and CrCl$_3$ were exfoliated onto SiO$_2$(285nm)/Si chip within a nitrogen-filled glove box (Inert Pure LabHE, $P_{O_2}$, $P_{H_2O}$ < 0.1ppm). By using a polymer-assisted pickup method reported previously[38], we sequentially stacked the structure of hBN/graphite/CrX$_3$/graphite/hBN in a home-built transfer setup inside the glove box, followed by transferring whole stack onto pre-patterned Au (40nm)/Ti (5nm) electrodes produced by conventional photolithography & lift-off methods and e-beam deposition. The overlapping area of graphite/CrX$_3$/graphite was set to be ~ 10 μm$^2$ for few-layer devices and ~1 μm$^2$ for bilayer devices. 1.4- to 7- nm-thick CrI$_3$ (2, 4, 8, and 10L), 1.3- to 7.8-nm-thick CrBr$_3$ (2, 8, 10, and 12L), and 1.2- to 9-nm-thick CrCl$_3$ (2, 8, 10, 12 and 15L) were used for fabrication. Graphite flakes were connected to pre-patterned electrodes and h-BN layers were used as protecting layers. Devices were stored inside the glovebox until they were loaded into the cryostat.

# Acknowledgments

AWT acknowledges support from the US Army Research Office (W911NF-19-10267), an Ontario Early Researcher Award (ER17-13-199), and the Korea-Canada Cooperation Program through the National Research Foundation of Korea (NRF) funded by the Ministry of Science, ICT and Future

Planning (NRF-2017K1A3A1A12073407). G-XM acknowledges support from an NSERC Discovery grant (RGPIN-04178). This research was undertaken thanks in part to funding from the Canada First Research Excellence Fund. HCL acknowledges support from the National Key R&D Program of China (Grants No. 2016YFA0300504), the National Natural Science Foundation of China (No. 11574394, 11774423, 11822412), the Fundamental Research Funds for the Central Universities, and the Research Funds of Renmin University of China (RUC) (15XNLQ07, 18XNLG14, 19XNLG17).

## Competing interests

The authors declare no competing interests.

## Supporting Information

The Supporting Information is available free of charge on the ACS Publications website.

    I-V characteristics for bilayer $CrX_3$ devices

    Determination of magnetic transition temperature

    Additional magnetotransport measurements for $CrCl_3$ and $CrBr_3$ devices

    Temperature-dependent TMR of $CrBr_3$ device

    Barrier height calculation

    Magnetotransport for $CrI_3$ device using $NbSe_2$ electrodes

**SYNOPSIS TOC**

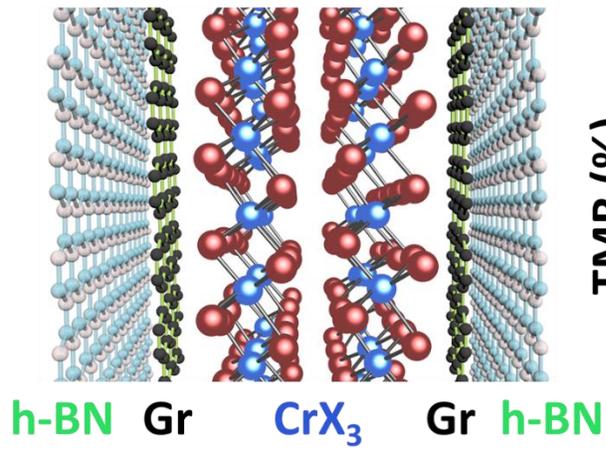 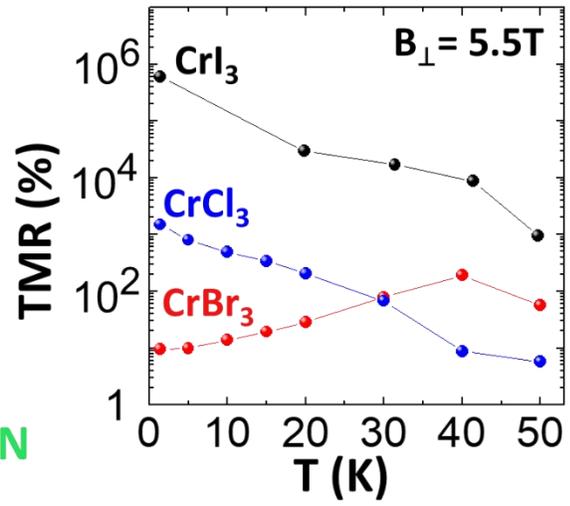

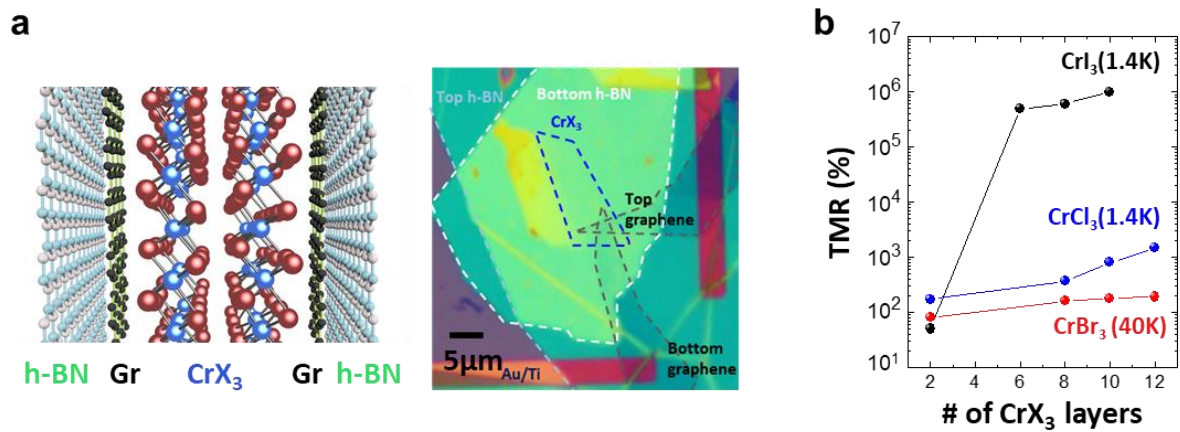

**Figure 1.** Van der Waals tunnel junctions incorporating ultrathin magnetic chromium trihalides. (a) Schematic illustration of the device (left) and an example optical image of a completed device with 2L $CrI_3$ (right). (b) Maximum TMR of $CrX_3$ as a function of the number of layers.

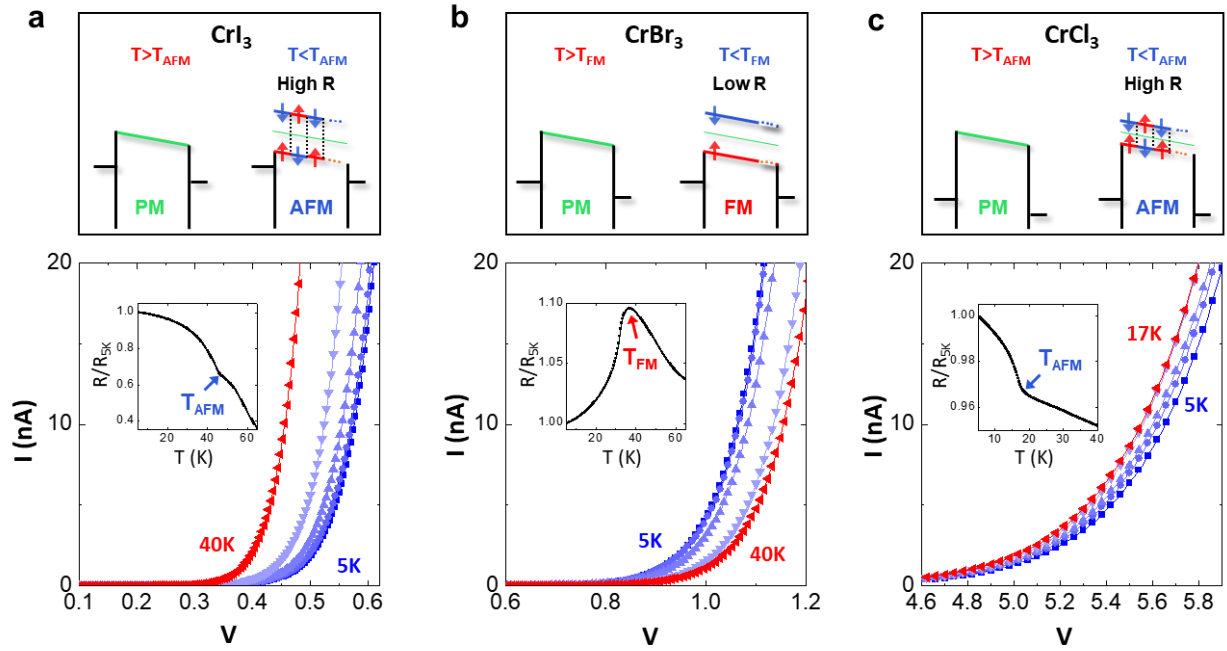

**Figure 2.** Temperature-dependent I-V measurement of (a) 8-layered $CrI_3$ (5, 10, 20, 30, and 40K), (b) 8-layered $CrBr_3$ (5, 10, 20, 30, and 40K), and (c) 15-layered $CrCl_3$ (5, 12, 14, 16, and 17K) at 0T. Insets show temperature-dependent, normalized d.c. resistance at 0T. Top panel shows respective, spin-dependent energy band diagrams of $Gr/CrX_3/Gr$ tunnel junctions above and below the magnetic transition temperature.

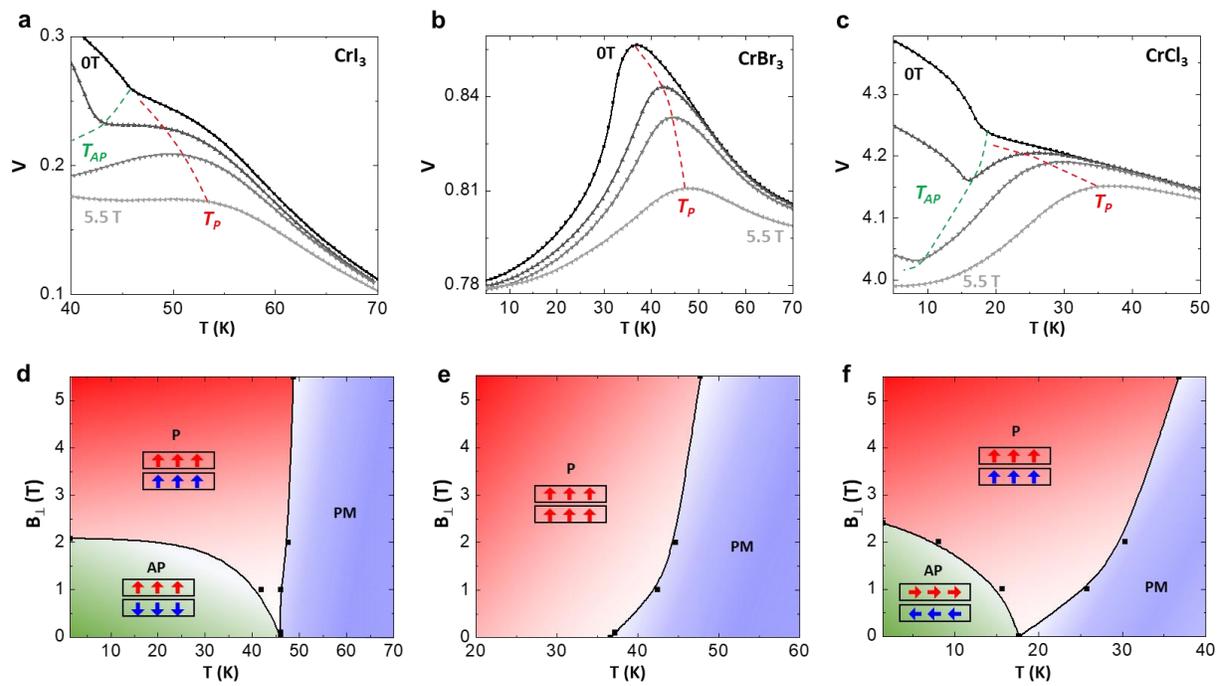

**Figure 3.** Tunneling probe of interlayer magnetic phases for three ultrathin chromium trihalides. (a) Voltage vs. temperature at 0.1 nA current biasing of a) 8-layer $CrI_3$, (b) 8-layer $CrBr_3$, and (c) 15-layer $CrCl_3$ for different $B_\perp$ (0, 1, 2, and 5.5 T, in sequence from top) (d-f) Field-temperature phase diagram obtained from (a-c).

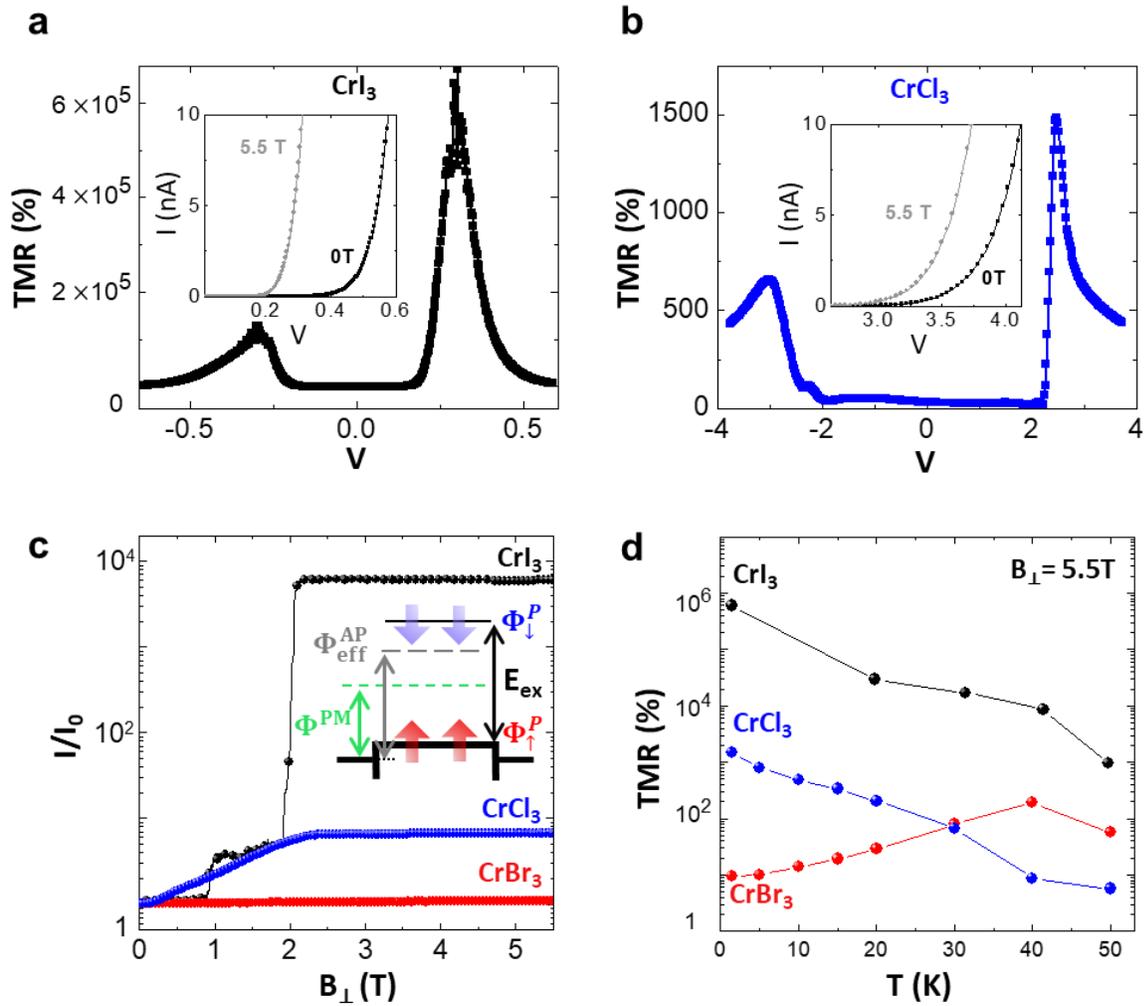

**Figure 4.** Tunnel magnetoresistance in three ultrathin chromium trihalides. TMR vs. voltage at 5.5T for in (a) 8L $CrI_3$ and (b) 12L $CrCl_3$ at 1.4K. Insets in (a) and (b) show full I-V characteristics at 0T and $B_\perp = 5.5T$ for the same devices. (c) Normalized tunnel current vs. perpendicular magnetic field at 1.4K for the same devices. Inset in (c) shows schematic illustration of spin-dependent energy band diagram for $CrX_3$ tunnel junction. (d) Temperature-dependent TMR at optimized voltage at $B_\perp = 5.5T$.

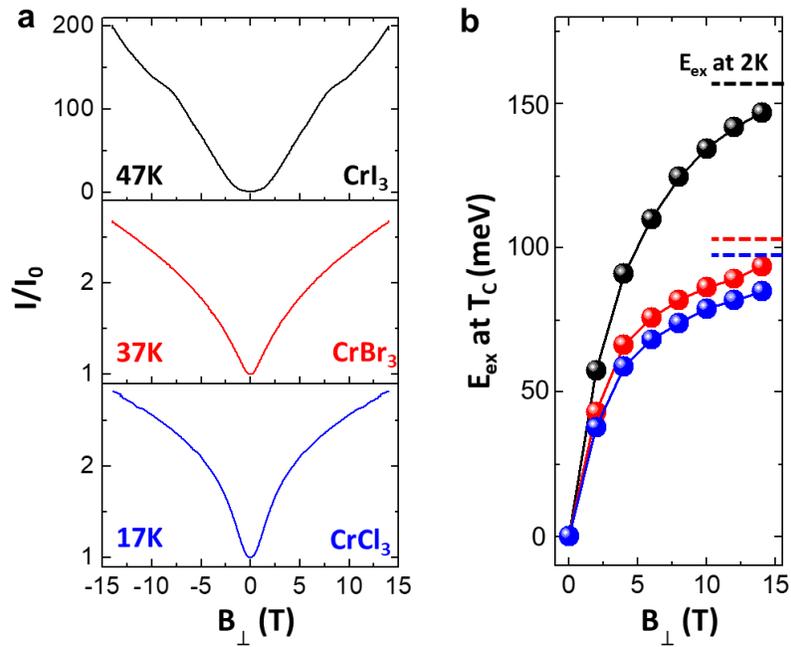

**Figure 5.** Non-saturating TMR behavior in few-layer $CrX_3$ at $T_C$. (a) field-dependent normalized current at optimized voltage (16L $CrI_3$: 0.33V, 10L $CrBr_3$: 1.3V, and 15L $CrCl_3$: 2.45V). (b) field-dependent exchange gap splitting for the same devices. Each dashed line indicates the spontaneous exchange splitting at 2K.

**Table 1.** Spin-dependent barrier height (in meV) between CrX$_3$ and few-layer graphene

|  | CrI$_3$ | CrBr$_3$ | CrCl$_3$ |
|---|---|---|---|
| $\Phi^{PM}$ | 227 | 538 | 943 |
| $\Phi^{AP}_{eff}$ @ 1.4K | 273 |  | 964 |
| $\Phi^{P}_{\uparrow}$ @ 1.4K | 159 | 477 | 909 |
| $\Phi^{P}_{\downarrow}$ @ 1.4K | 295 | 599 | 977 |

# Tailored tunnel magnetoresistance response in three ultrathin chromium trihalides

Hyun Ho Kim[1*], Bowen Yang[1], Shangjie Tian[2], Chenghe Li[2], Guo-Xing Miao[1], Hechang Lei[2], and Adam W. Tsen[1*]

[1]*Institute for Quantum Computing, Department of Chemistry, Department of Physics and Astronomy, and Department of Electrical and Computer Engineering, University of Waterloo, Waterloo, Ontario N2L 3G1, Canada*

[2]*Department of Physics and Beijing Key Laboratory of Opto-electronic Functional Materials & Micro-Nano Devices, Renmin University of China, Beijing 100872, China*

[*]Correspondence to: hyun.kim@uwaterloo.ca, awtsen@uwaterloo.ca

**This PDF file includes:**

I. I-V characteristics for bilayer $CrX_3$ devices

II. Determination of magnetic transition temperature

III. Additional magnetotransport measurements for $CrCl_3$ and $CrBr_3$ devices

IV. Temperature-dependent TMR of $CrBr_3$ device

V. Barrier height calculation

VI. Magnetotransport for $CrI_3$ device using $NbSe_2$ electrodes

## I. I-V characteristics for bilayer CrX$_3$ devices

In the main panel of Fig. S1, we show field-dependent I-V characteristics for three bilayer CrX$_3$ devices. The current is measurable down to zero bias. Maximum TMR in these samples is 50% for CrI$_3$, 80% for CrBr$_3$, and 170% for CrCl$_3$, all reduced from their thicker counterparts.

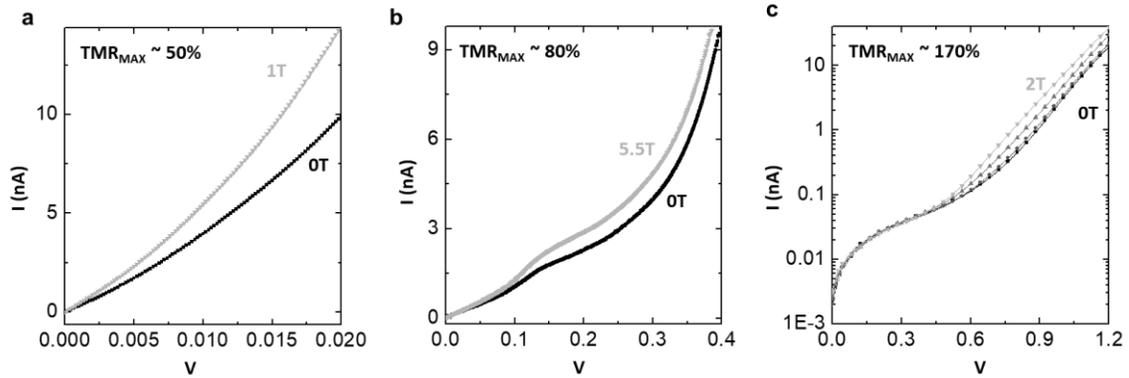

**Figure S1.** I-V characteristics for 2L (a) CrI$_3$, (b) CrBr$_3$, and (c) CrCl$_3$ devices. CrI$_3$ and CrCl$_3$ devices were measured at 1.4K and CrBr$_3$ device was measured at 40K.

## II. Determination of magnetic transition temperature

The magnetic transition temperature of $CrX_3$ can be more precisely determined by taking the first or second derivatives of voltage vs. temperature (see Fig. S2). At $T_{FM}$, $dV/dT = 0$ in $CrBr_3$, and at $T_{AFM}$, $d^2V/dT^2$ is peaked in $CrI_3$ and $CrCl_3$.

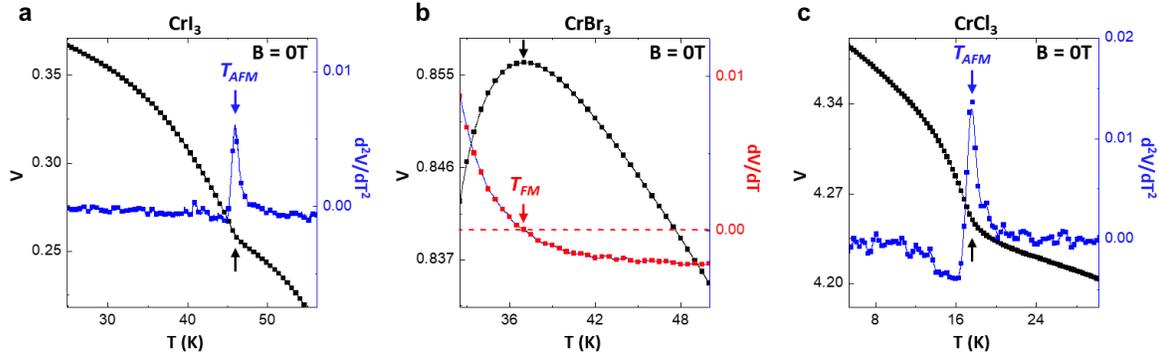

**Figure S2.** V vs. T plot and corresponding $dV/dT$ or $d^2V/dT^2$ of (a) 8L $CrI_3$, (b) 8L $CrBr_3$, and (c) 15L $CrCl_3$ at 0 T.

We measured voltage vs. temperature at several different currents for few-layer $CrI_3$ to determine the significance of Joule heating on the measured transition temperature (see Fig. S3). $T_{AFM}$ shows little change up to 100nA.

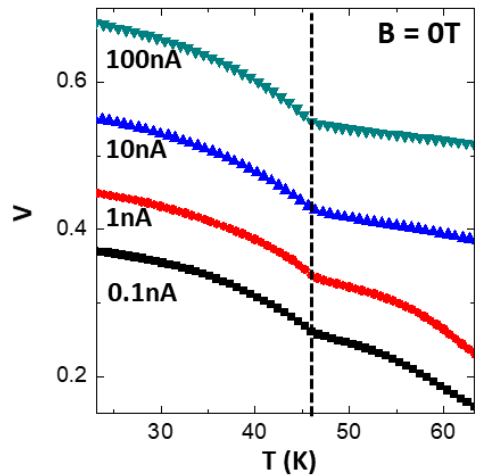

**Figure S3.** Voltage vs temperature of 8L $CrI_3$ for several different current biases.

### III. Additional magnetotransport measurements for CrCl$_3$ and CrBr$_3$ devices

In Fig. S4, we show additional voltage vs. temperature measurements at different perpendicular fields (0, 1, 2, 5.5T) for 10, 12L CrCl$_3$ devices and 10, 14L CrBr$_3$ devices. In Fig. S5, we show voltage vs. perpendicular magnetic field at several different temperatures for 8L CrBr$_3$ and 15L CrCl$_3$, the same devices shown in Fig. 3. The combined data set allow for the construction of the phase diagrams in Fig. 3d-f.

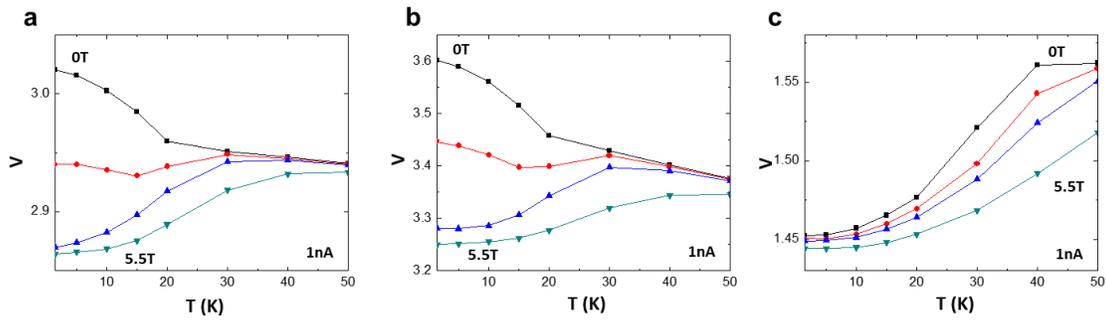

**Figure S4.** (a) Voltage vs temperature at 1nA for (a) 10L and (b) 12L CrCl$_3$ at $B_\perp = 0$, 1, 2, 5.5T. Same for (c) 10L CrBr$_3$.

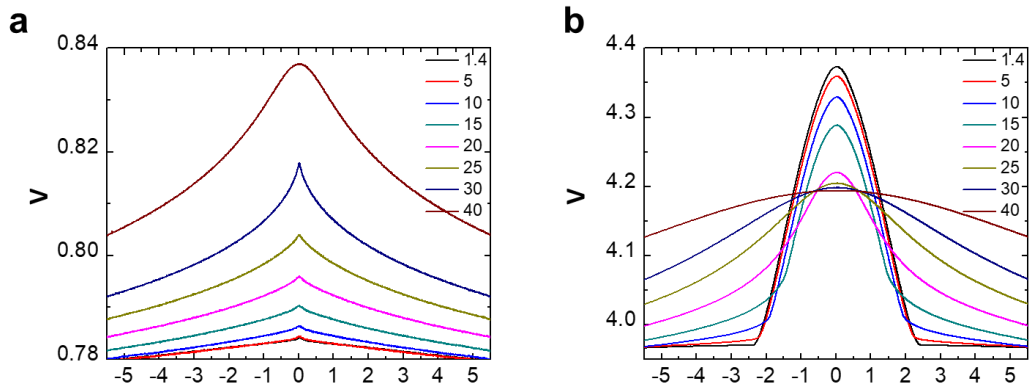

**Figure S5.** Voltage vs. perpendicular magnetic field at 1 nA for (a) 8L CrBr$_3$ and (b) 15L CrCl$_3$ at different temperatures.

## IV. Temperature-dependent TMR of CrBr₃ device

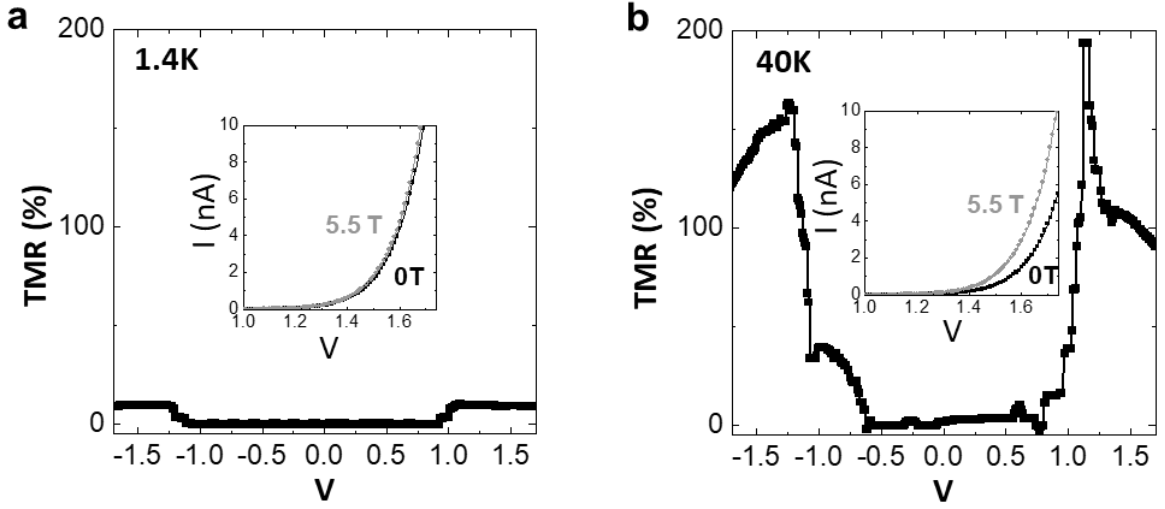

**Figure S6.** TMR vs. voltage at 5.5T for 12L CrBr3 at (a) 1.4K and (b) 40K. Insets in (a) and (b) show I-V characteristics at 0T and $B_\perp = 5.5$T for the same devices.

## V. Barrier height calculation

We calculated the spin-dependent barrier heights of CrX₃ devices using the Fowler-Nordheim (FN) current-voltage relation:

$$I \sim \frac{V^2}{\Phi} \exp\left(-\frac{4d\sqrt{2m\Phi^3}}{3\hbar eV}\right),$$

where $e$ is electronic charge, $d$ is the CrX₃ thickness, $m$ is the effective electron mass (estimated to be the free-electron mass), $\hbar$ is the Planck's constant, and $\Phi$ is the barrier height between the graphene electrode and CrX₃. We can linearize this relation by plotting $\ln\left(\frac{I}{V^2}\right)$ vs. $\frac{1}{V}$, and extracting $\Phi$ from the slope. This is shown in Fig. S7 for all three CrX₃. The various barrier heights are obtained under different field and temperature conditions.

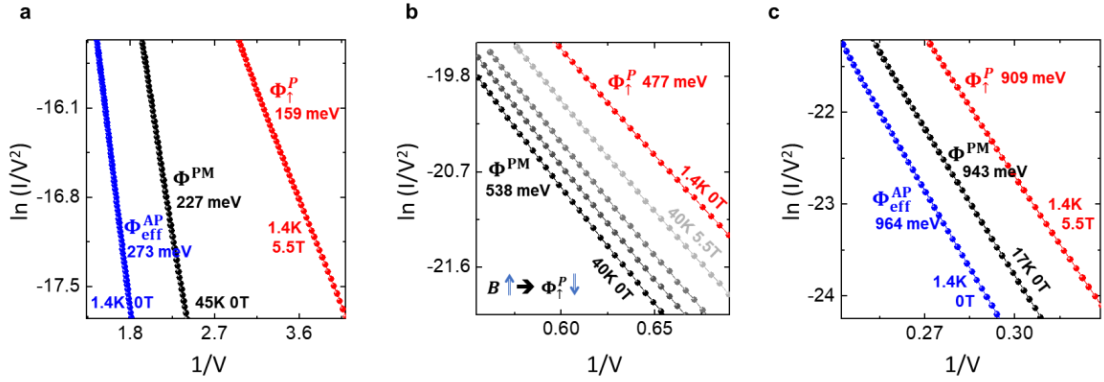

**Figure S7.** ln (I/V$^2$) vs.1/V for (a) 8L CrI$_3$, (b) 12L CrBr$_3$, and (c) 12L CrCl$_3$. Φ is determined by the slope. The magnetic field levels for the traces in (b) are 0, 1, 2, 5.5T from black to gray.

## VI. Magnetotransport for CrI$_3$ device using NbSe$_2$ electrodes

We additionally fabricated a CrI$_3$ device with high work function NbSe$_2$ (5.9eV) electrodes to investigate the spin-dependent current for hole transport (h-BN/NbSe$_2$/CrI$_3$/NbSe$_2$/h-BN). The results show that the critical polarizing field remains similar to that for devices using graphene electrodes, but with ~$10^4$ times lower TMR percentage, as shown in Fig. S8.

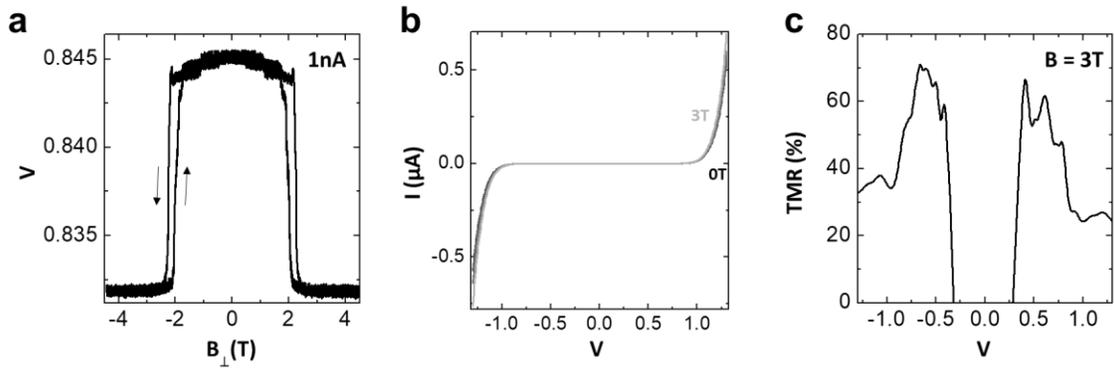

**Figure S8.** Magnetotransport results for NbSe$_2$/CrI$_3$/NbSe$_2$ tunnel junction device at 1.8K. The thickness for each material was measured to be ~10 nm. (a) Field-dependent voltage measurement at 1nA. (b) I-V plots from 0 to 3T, in sequence from black to grey. (c) voltage-dependent TMR at 3T.